\documentclass[english,aps, prb, twocolumn, superscriptaddress,showpacs,floatfix]{revtex4-1}
\usepackage{lmodern}
\usepackage[T1]{fontenc}
\usepackage[latin9]{inputenc}
\usepackage{textcomp}
\usepackage{graphicx}
\usepackage{bm}
\usepackage{verbatim}
\usepackage{booktabs}
\usepackage{amstext}
\usepackage{color}
\usepackage{upgreek}

\makeatletter

\newcommand{\lyxmathsym}[1]{\ifmmode\begingroup\def\b@ld{bold}
  \text{\ifx\math@version\b@ld\bfseries\fi#1}\endgroup\else#1\fi}

\providecommand{\tabularnewline}{\\}

\@ifundefined{textcolor}{}
{%
 \definecolor{BLACK}{gray}{0}
 \definecolor{WHITE}{gray}{1}
 \definecolor{RED}{rgb}{1,0,0}
 \definecolor{GREEN}{rgb}{0,1,0}
 \definecolor{BLUE}{rgb}{0,0,1}
 \definecolor{CYAN}{cmyk}{1,0,0,0}
 \definecolor{MAGENTA}{cmyk}{0,1,0,0}
 \definecolor{YELLOW}{cmyk}{0,0,1,0}
 \definecolor{darkgreen}{rgb}{0,0.6,0.4}
 }

\usepackage{prettyref}
\newrefformat{tab}{Tab.\,\ref{#1}}
\newrefformat{fig}{Fig.\,\ref{#1}}
\newrefformat{eq}{Eq.\,\textup{(\ref{#1})}}
\newcommand{\LDNFSxy}{Li$_{x}$(ND$_{2}$)$_{y}$(ND$_{3}$)$_{1-y}$Fe$_{2}$Se$_{2}$}

\newcommand{\LDNFS}{Li$_{0.6}$(ND$_{2}$)$_{0.2}$(ND$_{3}$)$_{0.8}$Fe$_{2}$Se$_{2}$}

\newcommand {\Tc} {\ensuremath{T_\mathrm{c}}}

\newcommand{\AFS}{$A_x$Fe$_{2-y}$Se$_2$}

\newcommand{\FeSeTe}{FeSe$_{1-x}$Te$_x$}

\newcommand{\spm}{s$_{\pm}$}

\makeatother

\usepackage{babel}

\begin{document}

\title{Spin fluctuations away from ($\pi$, 0) in the superconducting phase
of molecular-intercalated FeSe}

\author{A. E. Taylor }
\email[]{a.taylor1@physics.ox.ac.uk}

\affiliation{Department of Physics, University of Oxford, Clarendon Laboratory,
Parks Road, Oxford, OX1 3PU, United Kingdom}

\author{S. J. Sedlmaier}

\author{S. J. Cassidy}

\affiliation{Department of Chemistry, University of Oxford, Inorganic Chemistry Laboratory, South Parks
Road, Oxford, OX1 3QR, United Kingdom}

\author{E. A. Goremychkin}

\author{R. A. Ewings}

\author{T. G. Perring}

\affiliation{ISIS Facility, Rutherford Appleton Laboratory, STFC, Chilton, Didcot,
Oxon, OX11 0QX, United Kingdom}

\author{S. J. Clarke}

\affiliation{Department of Chemistry, University of Oxford, Inorganic Chemistry Laboratory, South Parks
Road, Oxford, OX1 3QR, United Kingdom}

\author{A. T. Boothroyd}
\email[]{a.boothroyd@physics.ox.ac.uk}

\affiliation{Department of Physics, University of Oxford, Clarendon Laboratory,
Parks Road, Oxford, OX1 3PU, United Kingdom}
\begin{abstract}
Magnetic fluctuations in the molecular-intercalated FeSe superconductor \LDNFSxy{} ($T_{\rm c} = 43$\,K) have been measured by inelastic neutron scattering from a powder sample. The strongest magnetic scattering is observed at a wave vector $Q \approx 1.4$ $\AA^{-1}$, which is not consistent with the $(\pi,0)$ nesting wave vector that characterizes magnetic fluctuations in several other iron-based superconductors, but is close to the $(\pi, \pi/2)$ position found for \AFS{} systems. At the energies probed ($\sim 5k_{\rm B}T_{\rm c}$), the magnetic scattering increases in intensity with decreasing temperature below \Tc{}, consistent with the superconductivity-induced magnetic resonance found in other iron-based superconductors.
\end{abstract}


\maketitle


The new molecular-intercalated FeSe compounds, with superconducting transition temperatures (\Tc{}) of up to 45\,K, present a new test-bed for understanding the Fe-based superconductors. \cite{ying_observation_2012,burrard-lucas_enhancement_2013} The inclusion of molecules between the FeSe layers, such as ammonia/amide and pyridine, in addition to alkali metal ions, appears to lengthen the $c$-axis and promote higher \Tc{}s than ever before seen in FeSe-based systems.\cite{ying_observation_2012,burrard-lucas_enhancement_2013,scheidt_superconductivity_2012,krzton-maziopa_synthesis_2012, ying_superconducting_2013}
An individual FeSe layer in these compounds is similar to a layer of pure FeSe, but the stacking of the layers along the $c$ axis is like in \AFS{} systems rather than in FeSe. So far the mechanism for the increased \Tc{} and its relationship with the \FeSeTe{} and \AFS{} superconductors remains unclear.

The maximum \Tc{} of the \FeSeTe{} series is $\sim$14.5\,K at ambient pressure,\cite{yeh_tellurium_2008,fang_superconductivity_2008}  rising to nearly 37\,K at pressures of 8.9\,kbar.\cite{medvedev_electronic_2009} 
Superconductivity has been found up to 30\,K in \AFS{} systems ($A=$ K, Rb, Cs).\cite{guo_superconductivity_2010, wang_superconductivity_2011, krzton-maziopa_synthesis_2011, fang_fe-based_2011} Unfortunately, \AFS{} superconductors are inhomogeneous and the precise composition of the superconducting phase is still under dispute, making the physics in these materials difficult to unravel.\cite{dagotto_unexpected_2012}

There is strong evidence that magnetic fluctuations couple to superconductivity in the Fe-based superconductors. Among the key observations is the so-called \emph{magnetic resonance peak}. This is a magnetic mode observed in neutron scattering spectra at an energy $E_{\rm res}$ close to the superconducting gap energy and at a well-defined wave vector ${\bf Q}_{\rm res}$, whose intensity increases on cooling through \Tc{}. The resonance behavior is usually ascribed to the BCS coherence factors.\cite{scalapino_common_2012}  For singlet pairing, the resonance peak results from strong scattering between portions of the Fermi surface connected by ${\bf Q}_{\rm res}$ on which the superconducting gap function has opposite sign.\cite{mazin_pairing_2009, hirschfeld_gap_2011} Therefore, measurements of the resonance peak can provide fundamental information about the superconducting state.

In common with many of the iron arsenide superconductors, the magnetic resonance peak of optimally-doped \FeSeTe{} is found at ${\bf Q}_{\rm res} = (\pi, 0)$ with respect to the Fe square lattice.\cite{dai_magnetism_2012, stewart_superconductivity_2011, lumsden_magnetism_2010, johnston_puzzle_2010} However, the \AFS{} systems have ${\bf Q}_{\rm res} = (\pi, \pi/2)$.\cite{park_magnetic_2011, taylor_spin-wave_2012, friemel_conformity_2012} The $(\pi, 0)$ wave vector corresponds to the displacement between quasi-nested hole and electron pockets on the Fermi surfaces of many iron-based superconductors, and the existence of a $(\pi, 0)$ resonance peak has been cited as strong evidence in favour of \spm{} symmetry of the superconducting gap.\cite{mazin_pairing_2009,maier_theory_2008,korshunov_theory_2008, yu_universal_2009, maier_neutron_2009} In contrast, the $(\pi, \pi/2)$ resonance peak, in conjunction with the lack of a hole pocket in ARPES measurements, was suggested to indicate $d$-wave pairing in \AFS{}.\cite{friemel_reciprocal-space_2012, maier_evolution_2012}

It is currently unclear where the molecular-intercalated FeSe systems fit into this picture. Yan and Gao performed Fermi surface calculations for alkali-metal-ion-intercalated FeSe, predicting different crystal structures and very different Fermi surfaces for \Tc{} $\sim 30$\,K and $\sim 40$\,K systems.\cite{yan_effect_2012} For the latter, they found the electronic structure to be very similar to that of the iron-arsenide systems. The \LDNFSxy{} system shows \Tc{}$\approx43\,$K, and a diffraction study determined its structure to be consistent with Yan and Gao's 40\,K model.\cite{burrard-lucas_enhancement_2013}  In addition, $\upmu$SR measurements on Li$_x$(C$_5$H$_5$N)$_y$Fe$_{2-z}$Se$_2$ found that the temperature dependence of the superconducting penetration depth is consistent with an \spm{} model.\cite{biswas_two-dimensional_2013}  These results seem to indicate that the molecular-intercalated FeSe systems are similar to \FeSeTe{}, and present different physics to \AFS{}. However, experiments with other techniques are needed to piece together a more complete picture of the superconductivity in these high-\Tc{} systems.


Here we present neutron inelastic scattering measurements on a molecular-intercalated FeSe system \LDNFSxy{}. We find strong magnetic fluctuations that increase on cooling below \Tc{}, consistent with a resonance peak. The  magnetic signal in momentum space is not described by the usual $(\pi, 0)$ wave vector, but is closer to $(\pi, \pi/2)$ as observed in \AFS{}. Our results suggest that \LDNFSxy{} could be similar to the minority superconducting phase found in \AFS{}.


The polycrystalline sample was prepared from tetragonal FeSe by the intercalation of lithium and ammonia between the layers via the route described in Ref.~\onlinecite{burrard-lucas_enhancement_2013}. Deuterated material was used to avoid a large incoherent scattering from protons in the neutron scattering experiments. The crystal structure and typical magnetization measurements are reported in Ref.~\onlinecite{burrard-lucas_enhancement_2013}. For a sample with \Tc{}$ =43\,$K, diffraction data revealed a composition \LDNFS{}, with lattice parameters $a=3.8059(1)\,$\AA{} and $c=16.1795(6)\,$\AA{} at 8\,K for the space group \emph{I}4/\emph{mmm}.

The inelastic neutron scattering experiments were performed on the
MERLIN time-of-flight chopper spectrometer at the ISIS Facility.\cite{bewley_merlin_2006}
The large, position-sensitive detector arrays on this instrument allow
us to search for magnetic excitations in a large region of $(Q,E)$
space in a single measurement. 11.4\,g of \LDNFSxy{}
powder was sealed inside a cylindrical aluminium can and mounted in
a top-loading closed-cycle refrigerator. All handling was carried out in an
inert gas atmosphere, and re-measurement of portions of the sample by SQUID magnetometry and X-ray and neutron diffraction confirmed that the samples were unchanged after the experiment.
Spectra were recorded with neutrons of incident energy $E{}_{\mathrm{{i}}}=80\,$ meV at a number of temperatures
between 5 and 67\,K. The energy resolution in this configuration
was \textasciitilde{}5.5\,meV, estimated from the full width at half
maximum of the incoherent part of the elastic peak.  The presented spectra have been normalised by the Bose population factor. The scattering from a standard vanadium sample was used to normalize the spectra
and place them on an absolute intensity scale, with units mb\,sr$^{-1}$\,meV$^{-1}$\,f.u.$^{-1}$,
where 1\,mb = 10$^{-31}\,$m$^{2}$ and f.u. stands for formula unit
of \LDNFS{}.

\begin{figure}
\includegraphics[clip,width=0.8\columnwidth]{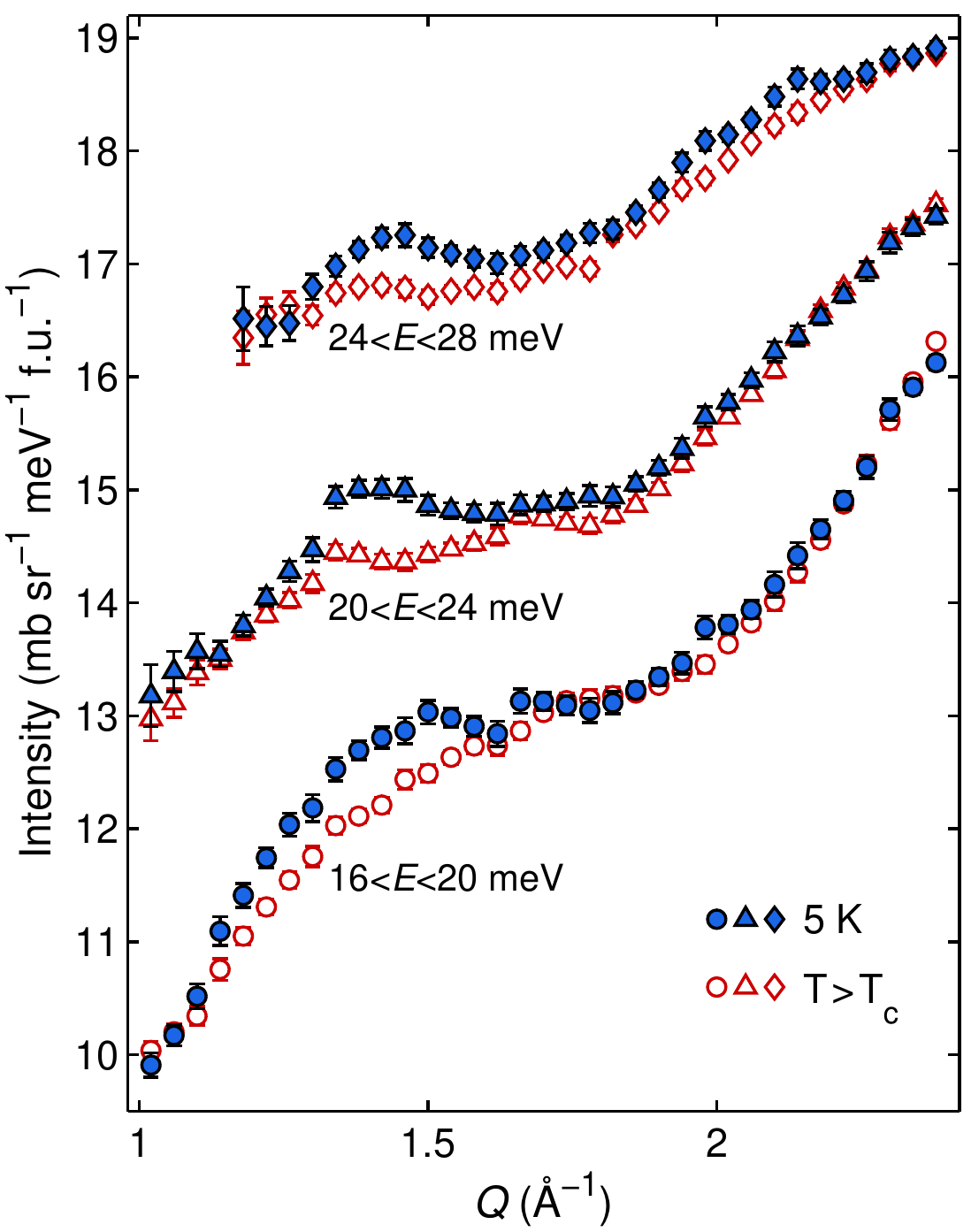}

\caption{\label{fig:Raw_data}(Color online) Neutron scattering from polycrystalline \LDNFSxy{} as a function of wave vector.  Data are shown averaged over three energy ranges as indicated, the upper two having being displaced vertically for clarity. The filled blue symbols represent data collected
at $5\,$K, and the open red symbols represent the $T>T_{\mathrm{{c}}}$
data, a combination of 58\,K and 67\,K data as described in the text. The intensities have been normalized
by the Bose factor $[1\lyxmathsym{\textminus}\exp(\lyxmathsym{\textminus}E/k_{\rm B}T)]^{\lyxmathsym{\textminus}1}$.}

\end{figure}
\begin{figure}
\includegraphics[width=0.5\columnwidth]{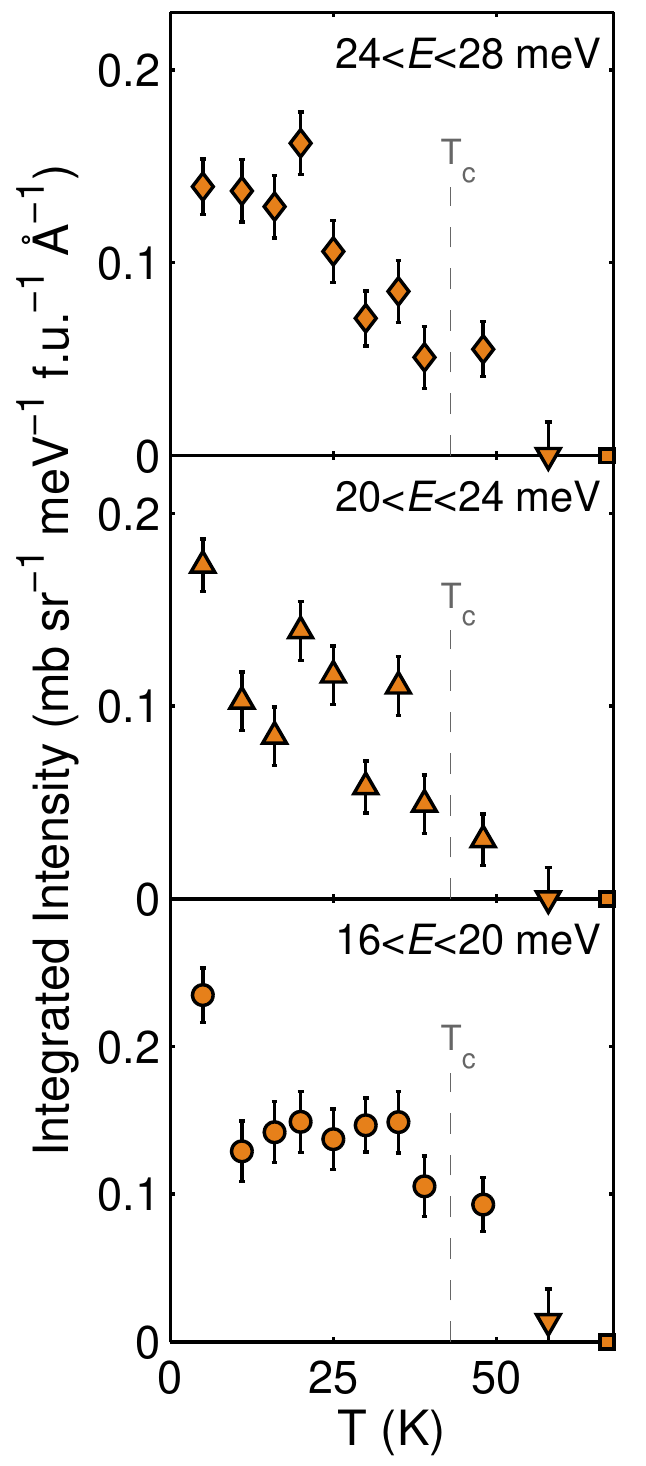}
\caption{(Color online) Integrated intensity of the signal at $Q\approx1.4\,\mathrm{{\AA}}^{-1}$ as a function of temperature for the three energy ranges indicated. Diamonds, upright triangles and circles all represent results of fits to $I(5\,{\rm K}) - I(T > T_{\rm c})$, where the $T>T_{\rm c}$ data is a combination of 58\,K and 67\,K data as described in the text. Inverted triangles are from similar fits to $I(58\,{\rm K}) - I(67\,{\rm K})$. Square symbols marks the zero reference point at $T=67\,{\rm K}$.
\label{fig:IIvsT}}

\end{figure}

Figure~\ref{fig:Raw_data} compares the scattering intensity from \LDNFSxy{} at temperatures above and below $T_\mathrm{c}$ for three energies between 16 and 28\,meV. Runs performed at 58 and 67$\,$K were used for the $T>T_\mathrm{c}$ reference data, and were combined in order to improve the statistics. The justification for averaging these runs is that after correction for the Bose population factor there was no detectable difference between the intensities measured at 58 and 67\,K (see Fig.~\ref{fig:IIvsT}).

All three constant-energy cuts shown in Fig.~\ref{fig:Raw_data} exhibit a significant difference between the response at 5\,K and at $T>T_\mathrm{c}$. We expect the scattering intensity at these energies to be due to phonon and inelastic magnetic scattering processes, with phonon scattering accounting for the general increase in signal with $Q$ seen in Fig.~\ref{fig:Raw_data}. However, within the $(Q,E)$ region shown we can reasonably expect the Bose factor correction to nullify the change in phonon scattering intensity with temperature, so we attribute the extra intensity at 5\,K to magnetic scattering.

\begin{figure}
\includegraphics[width=0.8\columnwidth]{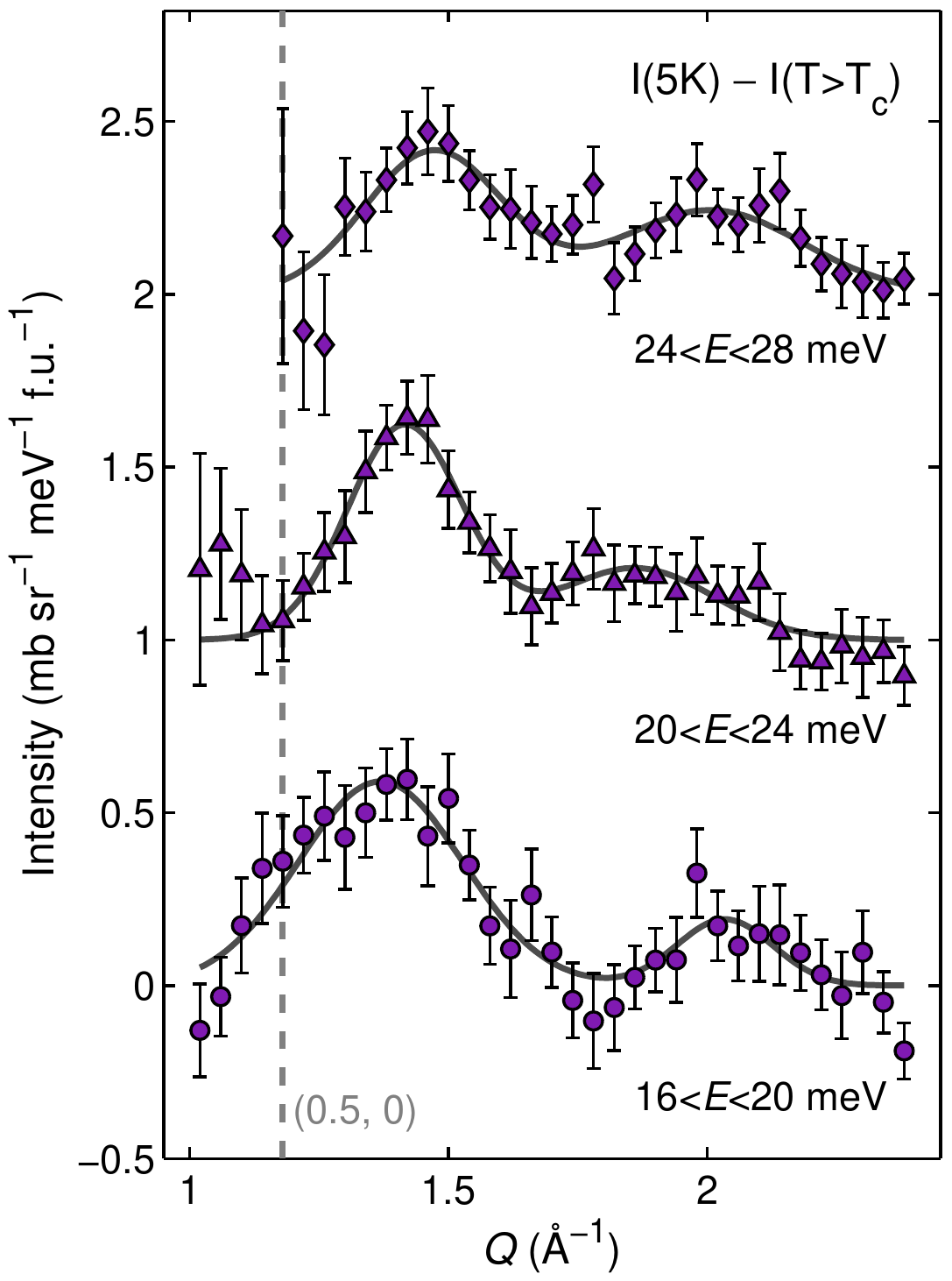}

\caption{\label{fig:Difference_Plots}(Color online) Difference between the intensity measured at $T<T_{\mathrm{{c}}}$ and $T>T_{\mathrm{{c}}}$ for each pair of constant-energy cuts shown in Fig.~\ref{fig:Raw_data}. Successive plots are displaced vertically
by one unit for clarity. The solid lines are the results of fits to two
Gaussian peaks, as described in the text. The wave vector corresponding to the position $(0.5,0,0) \equiv (\pi, 0, 0)$ in momentum space is marked by the dashed line for reference. }

\end{figure}

A clearer picture of the magnetic scattering is provided by Fig.~\ref{fig:Difference_Plots}, which displays the difference between the intensity at 5\,K and at $T>T_\mathrm{c}$. Each cut contains two peaks, one centered at $Q_1\approx 1.4\,\mathrm{{\AA}}^{-1}$ and the other at $Q_2\approx 2\,\mathrm{{\AA}}^{-1}$. To quantify these peaks we fitted the subtracted data to two Gaussian functions, allowing the width, center, and amplitude of each Gaussian to vary independently. The fitted centers ($Q_{i}$) and widths ($\sigma_{i}$) are given in Table~\ref{tab:Cen_sigmas}. In subsequent fits at other temperatures (not shown) the peak centers and widths were constrained to the values in Table~\ref{tab:Cen_sigmas} and only the areas of the peaks were allowed to vary.

%
\begin{table}
\begin{tabular}{ccccc}
\toprule
Energy (meV) & $Q_{1}$ ($\mathrm{{\AA}}^{-1}$) & $\sigma_{1}$ ($\mathrm{{\AA}}^{-1}$) & $Q_{2}$ ($\mathrm{{\AA}}^{-1}$) & $\sigma_{2}$ ($\mathrm{{\AA}}^{-1}$)\tabularnewline
\midrule
$16<E<20$ & 1.37(2) & 0.16(2) & 2.03(4) & 0.09(4)\tabularnewline
$20<E<24$ & 1.42(1) & 0.11(1) & 1.86(5) & 0.15(5)\tabularnewline
$24<E<28$ & 1.47(2) & 0.13(3) & 2.01(4) & 0.18(5)\tabularnewline
\bottomrule
\end{tabular}
\caption{Results of fitting two Gaussian functions to the data shown in Fig.~\ref{fig:Difference_Plots}. The best-fit parameters and errors (in parentheses) are the result of a least-squares fitting procedure. The $Q_i$ are the Gaussian peak centers and the $\sigma_i$ are the corresponding standard deviations, where $\sigma = \mathrm{FWHM}/(2\sqrt{2\ln2})$.  \label{tab:Cen_sigmas}}
\end{table}

The peak area (integrated intensity) gives a measure of the strength of the magnetic fluctuations. The areas of the fitted $Q_1\approx 1.4\,\mathrm{{\AA}}^{-1}$ peaks are plotted in Fig.~\ref{fig:IIvsT} as a function of temperature. There is a general trend of increasing area with decreasing temperature below \Tc{}. The data are not of sufficient statistical quality to extract a meaningful trend for the area of the $Q_2\approx 2\,\mathrm{{\AA}}^{-1}$ peak as a function of temperature, however this peak was included in all fits to avoid attributing excess signal to the lower $Q$ peak.

\begin{figure}
\includegraphics[width=0.9\columnwidth]{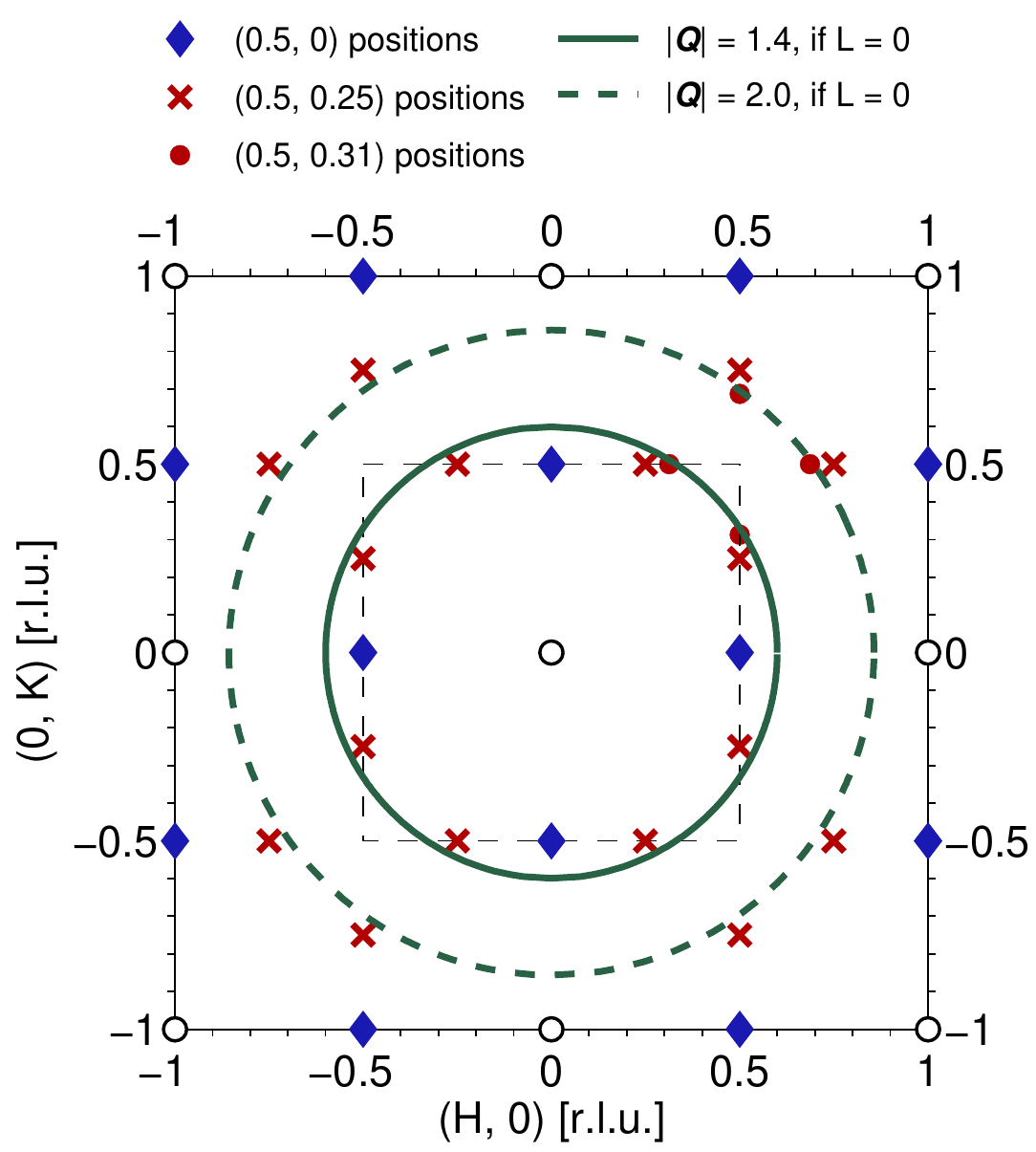}\caption{(Color online) Map of two-dimensional reciprocal space for \LDNFSxy{} referred to the one-Fe unit cell. The dashed square marks the first Brillouin zone boundary. The solid and dashed rings show the values of $Q$ where magnetic signals are observed in our powder data, if that signal is assumed to have no out-of-plane component. The additional symbols in the upper-right quadrant show the first and second order resonance peak positions predicted in Ref.~\onlinecite{Maier_d-wave_2011}.
\label{fig:Recip_Map}}

\end{figure}

To interpret the results we need to relate the  powder-averaged  $Q$ values of the magnetic peaks to wave vectors in the Brillouin zone.
Figure~\ref{fig:Recip_Map} is a map of the $(H,K)$ plane in two-dimensional (2D) reciprocal space. We neglect the out-of-plane wave vector component for now, and we index positions with respect to the one-Fe unit cell which has in-plane lattice parameter $a=b=2.691\,$\AA. The map shows the positions of previous observations of a neutron spin resonance in iron-based superconductors at $(0.5,0)$  $[ \equiv (\pi,0)]$ and $(0.5,0.25)$  $[ \equiv (\pi,\pi/2)]$.  The circles represent the locus of points in the 2D Brillouin zone that have $Q=1.4\,\mathrm{{\AA}}^{-1}$ and $2.0\,\mathrm{{\AA}}^{-1}$, corresponding to the two peak positions in Fig.~\ref{fig:Difference_Plots}.


%

It is immediately clear from Fig.~\ref{fig:Recip_Map} that the wave vector $(0.5,0)$ and equivalent positions cannot account for the $Q_1$ and $Q_2$ values at which we observe magnetic scattering. The wave vector corresponding to  $(0.5,0)$ is also marked on Fig.~\ref{fig:Difference_Plots} to show that it is displaced away from the maximum of the $Q_1$ peak. We also find no evidence for magnetic scattering at wave vectors such as $(0.7, 0.1)$ ($Q = 1.65\,\mathrm{{\AA}}^{-1}$) where magnetic order and strong magnetic fluctuations are observed in the $\sqrt{5}\times\sqrt{5}$ Fe vacancy-ordered phase of the bi-phasic \AFS{} superconductors. We do, however, find that the circles of radius  $Q_1$ and $Q_2$ pass quite close to the $(0.5, 0.25) \equiv (\pi, \pi/2)$ set of wave vectors and their second order positions $(0.5, 0.75) \equiv (\pi, 3\pi/2)$, etc., where the resonance is seen in  \AFS{} superconductors.\cite{park_magnetic_2011, taylor_spin-wave_2012, friemel_conformity_2012}

We now consider the effect of the out-of-plane wave vector component, $L$, on the peak positions. The magnetic fluctuations are likely to be 2D like those in \FeSeTe{} and \AFS{},\cite{qiu_spin_2009, friemel_reciprocal-space_2012} therefore we expect the magnetic signal to be highly extended in the $(0,0,L)$ direction. The effect of powder-averaging on 2D scattering is to shift the peak to a higher $Q$ than $Q = |(H, K, 0)|$  due to the contribution from $(H, K, L\neq0)$ (which, however, diminishes with increasing $L$ due to the magnetic form factor of Fe). We can estimate this shift from inelastic neutron scattering measurements on powder and single crystal samples of LiFeAs.\cite{taylor_antiferromagnetic_2011, qureshi_inelastic_2012} The magnetic peak in the powder data is at $Q = 1.24\,{\rm \AA}^{-1}$, whereas the observed in-plane wave vector $(0.5, \pm 0.07)$ has magnitude $Q = 1.19\,{\rm \AA}^{-1}$, giving a shift due to powder-averaging of $\Delta Q  = 0.05\,{\rm \AA}^{-1}$. Applying this correction to the resonance wave vector of \AFS{} we obtain $Q_{\rm res} = |(0.5, 0.25, 0)| + \Delta Q = 1.36\,{\rm \AA}^{-1}$, which is close to, but smaller than  $Q_1 = 1.4\,{\rm \AA}^{-1}$ observed here.

This analysis suggests that the peak at $Q_{1}$ cannot be explained simply by the effect of powder averaging a 2D signal with wave vector $(0.5, 0.25, L)$.  This conclusion is supported by the fact that $Q_{2} \approx 2.0\,{\rm \AA}^{-1}$ is lower than the value expected from $|(0.5, 0.75, 0)| = 2.10\,{\rm \AA}^{-1}$ (Fig.~\ref{fig:Recip_Map}). Interestingly, however, the wave vector ${\bf Q}_{\rm res} = (0.5, 0.31)$ predicted from band structure calculations of \AFS{} (Ref.~\onlinecite{Maier_d-wave_2011}) reproduces both the $Q_1$ and $Q_2$ peaks very well, as shown in Fig.~\ref{fig:Recip_Map}.

 The temperature dependence of the magnetic peak, Fig.~\ref{fig:IIvsT}, is similar to that of resonance peaks observed in other Fe-based superconductors, with an increase in the intensity with decreasing temperature below \Tc{} (or starting slightly above \Tc{}). This behavior is often cited as evidence for a link between magnetic fluctuations and superconductivity (for a review see Ref.~\onlinecite{li_superconductivity_2011}).

The lowest temperature point of the $16<E<20$\,meV data in Fig.~\ref{fig:IIvsT} has an anomalously high integrated intensity, which correlates with an anomalously large peak width --- see Table~\ref{tab:Cen_sigmas}. Inspection of Figs.~\ref{fig:Raw_data} and~\ref{fig:Difference_Plots} shows that this increased width appears to be caused by additional intensity on the low $Q$ side of the peak. The origin of this additional scattering is not known, but one possibility is the presence of a magnetic resonance mode with a wave vector near $(0.5,0)$. This could originate from a secondary superconducting phase with a \Tc{} of between 5 and 10\,K. An impurity of tetragonal FeSe would be a potential secondary phase in \LDNFSxy{},
but X-ray and neutron diffraction measurements on the sample used in this experiment rules out FeSe above the 4\,wt\% level. It is also possible that the anomalous intensity is related to the increase in relaxation below $\sim 10$\,K observed in the \textmu{}SR measurements  on the same material.\cite{burrard-lucas_enhancement_2013}


The resonance peak in other Fe-based superconductors is observed over a limited range of energy around $E_{\rm res} \sim 5k_{\rm B}$\Tc{}. For the sample of \LDNFSxy{} studied here, $5k_{\rm B}$\Tc{} $\approx 19$\,meV, so the enhancement in intensity observed below \Tc{} in Fig.~\ref{fig:IIvsT} is consistent with a magnetic resonance with $E_{\rm res} \sim 5k_{\rm B}$\Tc{}. However, to confirm this it is desirable to extend measurements of the spectrum to higher and lower energies than we could probe in this experiment.\cite{footnote}

Since our intensity measurements are calibrated we can also compare the strength of the magnetic signal found here to that observed for other Fe-based superconductors. The integrated intensity for the $24<E<28\,$meV $Q$-cut at 5\,K (Fig.~\ref{fig:IIvsT}) is 0.07(1)\,mb\,sr$^{-1}$\,meV$^{-1}$\,\AA$^{-1}$ per Fe (the formula unit contains two Fe atoms). A similar powder measurement on superconducting LiFeAs, Ref.~\onlinecite{taylor_antiferromagnetic_2011}, found the integrated intensity at the peak energy of the magnetic resonance to be 0.073(5)\,mb\,sr$^{-1}$\,meV$^{-1}$\,\AA$^{-1}$ per Fe at 6\,K, which is known to be similar in strength to that found in other Fe-based superconductors,\cite{taylor_antiferromagnetic_2011, wang_antiferromagnetic_2011} including \AFS{}.\cite{taylor_spin-wave_2012, friemel_conformity_2012} Therefore, the magnetic signal we have observed in \LDNFSxy{} is consistent in strength with the resonance peaks in other Fe-based superconductors.

The observation of resonance-like magnetic peaks is not unexpected, but their positions at $Q_1$ and $Q_2$ away from $|(0.5, 0)|$  (see Fig. \ref{fig:Recip_Map}) is surprising given the results of $\upmu$SR\cite{biswas_two-dimensional_2013} and Fermi surface calculations\cite{yan_effect_2012} which suggest that these materials are similar to \FeSeTe{} and iron-arsenide superconductors. It is also intriguing that, despite similar temperature dependence, energy scale and absolute intensity, the signal is also not fully explained by the ${\bf Q}_{\rm res} = (0.5, 0.25) \equiv (\pi, \pi/2)$  as observed for \AFS{}, but is very close to an initial prediction made from a band structure calculation for \AFS{} (Ref.~\onlinecite{Maier_d-wave_2011}). 

In conclusion, we have observed magnetic fluctuations in \LDNFSxy{} consistent with a superconductivity-induced resonance peak at wave vectors that are distinct from the $(\pi,0)$ nesting wave vector that characterizes magnetic fluctuations in \FeSeTe{}. The magnetic wave vectors are better matched to those of the superconducting component of \AFS{}, although the match is not perfect. We find no evidence of a signal at the wave vector corresponding to the $\sqrt{5}\times\sqrt{5}$ magnetically ordered component of \AFS{}.  Since the position of the magnetic resonance has important implications for the symmetry of the pairing function, these results provide the motivation for better band structure calculations and theory to understand the nature of superconductivity in this material.


Acknowledgements: This work was supported by the UK Engineering \&
Physical Sciences Research Council and the Science \& Technology Facilities Council. We thank Peter Hirschfeld for helpful discussions.


\bibliographystyle{apsrev4-1}
\bibliography{Intercal_bib}

\end{document}